\documentclass{aa}
\usepackage[varg]{txfonts}

\usepackage{natbib}
\bibpunct{(}{)}{;}{a}{}{,}
\usepackage{array}
\usepackage{xspace}
\usepackage{lscape}
\usepackage{multirow}
\usepackage{url}
\usepackage[normalem]{ulem}
\usepackage{amsmath}
\usepackage{color}
\usepackage{calrsfs}
\usepackage[bottom]{footmisc}
\usepackage[hyperfootnotes=false, linktocpage=true, breaklinks=true, colorlinks=true, linkcolor=blue, citecolor=blue, urlcolor=blue]{hyperref}
\usepackage[all]{hypcap}
\usepackage{graphicx}

\newcommand{\NH}{\ensuremath{N_{\mathrm{H}}}\xspace}

\newcommand{\slab}{{\tt slab}\xspace}

\newcommand{\amol}{{\tt amol}\xspace}
\newcommand{\hot}{{\tt hot}\xspace}
\newcommand{\pow}{{\tt pow}\xspace}
\newcommand{\bb}{{\tt bb}\xspace}

\newcommand{\combinespectra}{{\tt combine\_grating\_spectra}\xspace}
\newcommand{\sectors}{{\tt sectors}\xspace}

\newcommand{\xmm}{{\rm XMM$-$\it Newton}\xspace}
\newcommand{\chandra}{{\it Chandra}\xspace}
\newcommand{\spex}{{SPEX}\xspace}

\newcommand{\La}{\mathcal{L}}

\newcommand{\ovii}{\ion{O}{vii}\xspace}
\newcommand{\oviii}{\ion{O}{viii}\xspace}

\mathchardef\mhyphen="2D

\mathchardef\mhyphen="2D

\begin{document}


\title{Oxygen and iron in interstellar dust: an X-ray investigation}

\author{
I. Psaradaki \inst{1,2}
\and
E. Costantini \inst{1,3}
\and
D. Rogantini \inst{8,1}
\and 
M. Mehdipour \inst{9}
\and
L. Corrales \inst{2}
\and 
S.T. Zeegers \inst{7}
\and
F. de Groot \inst{4}
\and 
J.W.A. den Herder \inst{1,3}
\and
H. Mutschke \inst{5}
\and
S. Trasobares \inst{6}
\and 
C. P. de Vries  \inst{1}
\and
L.B.F.M. Waters \inst{1,10}
}

\institute{
SRON Netherlands Institute for Space Research, Niels Bohrweg 4, 2333 CA Leiden, the Netherlands
\and
University of Michigan, Dept. of Astronomy, 1085 S University Ave, Ann Arbor, MI 48109, USA
\and
Anton Pannekoek Astronomical Institute, University of Amsterdam, P.O. Box 94249, 1090 GE Amsterdam, the Netherlands
\and 
Debye Institute for Nanomaterials Science, Utrecht University, Universiteitsweg 99, 3584 CG Utrecht, Netherlands
\and 
Astrophysikalisches Institut und Universitats-Sternw\"arte (AIU), Schillerg\"a{\ss}chen 2-3, 07745 Jena, Germany
\and
Departamento de Ciencia de los Materiales e Ingenier\'ia Metal\'urgica y Qu\'imica Inorg\'anica, Facultad de Ciencias, Universidad de C\'adiz, Campus R\'io San Pedro, Puerto Real, 11510, C\'adiz, Spain
\and 
Academia Sinica, Institute of Astronomy and Astrophysics, 11F Astronomy-Mathematics Building, NTU/AS campus, No. 1, Section 4, Roosevelt Rd., Taipei 10617, Taiwan
\and 
MIT Kavli Institute for Astrophysics and Space Research, Cambridge, MA 02139, USA
\and 
Space Telescope Science Institute, 3700 San Martin Dr, Baltimore, MD 21218, USA
\and
Department of Astrophysics/IMAPP, Radboud University P.O. Box 9010, 6500 GL Nijmegen, the Netherlands}



\abstract
{Understanding the chemistry of the interstellar medium (ISM) is fundamental for the comprehension of the Galactic and stellar evolution. X-rays provide an excellent way to study the dust chemical composition and crystallinity along different sight-lines in the Galaxy. In this work we study the dust grain chemistry in the diffuse regions of the interstellar medium in the soft X-ray band (<1 keV). We use newly calculated X-ray dust extinction cross sections, obtained from laboratory data, in order to investigate the oxygen K and iron L shell absorption. We explore the \xmm and \chandra spectra of 5 low-mass X-ray binaries located in the Galactic plane, and we model the gas and dust features of oxygen and iron simultaneously. The dust samples used for this study include silicates with different Mg:Fe ratios, sulfides, iron oxides and metallic iron. Most dust samples are in both amorphous and crystalline lattice configuration. The extinction cross sections have been computed using Mie scattering approximation and assuming a power law dust size distribution. 
We find that the Mg-rich amorphous pyroxene ($\rm Mg_{0.75}Fe_{0.25}SiO_{3}$) represents the largest fraction of dust towards most of the X-ray sources, about 70\% on average. Additionally, we find that $\rm \sim$ 15\% of the dust column density in our lines of sight is in Fe metallic. We do not find strong evidence for ferromagnetic compounds, such as $\rm Fe_{3}O_{4}$ or iron sulfides ($\rm FeS$, $\rm FeS_{2}$). Our study confirms that the iron is heavily depleted from the gas phase into solids; more than 90\% of iron is in dust. The depletion of neutral oxygen is mild, between 10-20\% depending on the line of sight.}

\authorrunning{I. Psaradaki et al.}
\titlerunning{Oxygen and iron in interstellar dust: an X-ray investigation}
\maketitle


\section{Introduction}

Interstellar dust is ubiquitous and plays an important role in our Galaxy. Dust appears at every stage of stellar evolution, from evolved stars and supernovae to proto-planetary discs. Dust as the primary repository of metals in the interstellar medium (ISM), can regulate its thermal structure and provide the surface for chemical reactions \citep{Henning2010}. The properties of dust grains in different regions of the ISM can give insights into their production and destruction mechanisms and reveal the evolution history of our Galaxy.

Dust grains are characterized by their chemical composition, morphology and size. Abundant elements such as carbon, oxygen, silicon, iron and magnesium are the main constituents of cosmic dust. Elements such as titanium, calcium and aluminum can be found in smaller quantities and are highly depleted from the gas phase (\citealt{Jenkins2009}). Cosmic dust can roughly be divided into \textit{carbonaceous} and \textit{silicate} grains. Additionally, dust can consist of oxides (e.g. $\rm Fe_{2}O_{3}$), sulfides (e.g. $\rm FeS$), carbides and iron inclusions, such as metallic iron (\citealt{Drainebook}).

 The structural properties of dust grains vary depending on the configuration of the atoms in the grain. \textit{Crystalline} materials are characterized by a periodic long-range order of atoms, while \textit{amorphous} dust grains do not show periodic structures and present a 3D disordered network of atoms. Crystalline dust is unlikely to survive the harsh environment of the ISM. Studies of the 10 $\mu$m silicate feature in the mid-IR band showed that interstellar silicate dust is mostly in amorphous form, and in particular olivines and Mg-rich pyroxenes (\citealt{Kemper2004}, \citealt{Min2007}). Moreover, it has been recently discovered from X-ray absorption studies in the Si K and Mg K spectral regions, that amorphous olivine is a dominant chemical composition in the dense parts of the diffuse ISM (\citealt{Zeegers2017, Zeegers2019}, \citealt{Rogantini2019, Rogantini2020}). However, the exact ratio of olivines to pyroxenes in dust along with the Mg:Fe ratio in silicates is not yet fully constrained. This is an important issue to resolve as Mg-rich silicates (rather than Fe-rich) could give an explanation for the high amount of Mg found in cometary and circumstellar grains (\citealt{Min2007}). At the same time, this constrains the exact amount of iron in silicates which is still unknown.

Historically, interstellar dust has been studied in the infrared and millimeter wavelengths. However, the launch of the \xmm and \chandra X-ray satellites and their high-resolution spectrometers opened up a new window for the study of the ISM and the dust mineralogy. The photoabsorption edges of some of the most abundant elements in the Galaxy, such as oxygen and iron are present in the X-ray spectra of many of the brightest Galactic sources (\citealt{Schattenburg1986}, \citealt{Paerels2001}, \citealt{Ueda2005}). Today, bright X-ray binaries are used as background lights for studying the spectral features of atomic and solid species of the ISM. In particular, with high-resolution X-ray spectra we are able to distinguish the gas and solid phase abundances of individual elements as well as to investigate the chemical composition of dust in the ISM (e.g. \citealt{Wilms}, \citealt{Costantini2012}, \citealt{Pinto2013} and references therein).

Oscillatory modulations, known as X-ray Absorption Fine Structures (XAFS), are observed near the photoelectric absorption edges, and they provide unique fingerprints of dust. These modulations happen when an X-ray photon gets absorbed by an atom in the dust grain. The ejected photo-electron interacts with the neighboring atoms. This interaction modulates the wave-function of the photo-electron due to constructive and destructive interference between the outgoing and backscattered electron waves. The shape of the spectral modulations is determined by the local-scale atomic structure of the grain which imprints its chemical composition and structure \citep[see also][for a detailed explanation]{Newville}. Therefore, modelling the XAFS enables the study of the dust chemical composition and crystallinity.

Up-to-date models of XAFS, based on laboratory experiments, have been recently released in the \amol model of the \spex X-ray fitting package\footnote{\url{https://spex-xray.github.io/spex-help/models/amol.html}}. In particular, the iron K edge has been studied by \citet{Rogantini2018} using synchrotron radiation. Synchrotron measurements have been also performed in order to characterize the K edges of silicon (\citealt{Zeegers2017,Zeegers2019}) and magnesium (\citealt{Rogantini2019}). For the soft X-ray band, the oxygen K edge (\citealt{Psaradaki}, henceforth, Paper I) and the iron L edges (Costantini et al. in prep.) have been studied using laboratory measurement from an electron energy loss spectrometer (\citealt{Egerton}). Additionally, studies of the Fe L edges have been performed in the past by \citet{Lee2009b} and \citet{Westphal} and for the Fe K edge by \citet{Lee2005}. Elements such as carbon or of lesser abundance in the ISM such as aluminum, calcium and sulfur have been presented in \citet{Costantini2019}.

In this work we focus on iron and oxygen content of interstellar dust. The question "Where is the interstellar iron?" is still a long-standing conundrum. Iron is known to be predominantly depleted from the gas phase into solids, with the depletion value changing slightly depending on the environment. Despite this high depletion rate of Fe, the exact composition of the Fe bearing grains is still unclear. Iron is primarily formed in supernovae type Ia and core collapse supernovae. It is believed that more than $\rm 65 \% $ of the iron is injected in the ISM in the gaseous form and therefore most of its grain growth is expected to take place in the ISM (\citealt{Dwek2016}).

\citet{Schalen1965} and \citet{Kemper2002} suggested that some of the depleted cosmic iron could be in the form of pure metallic iron. \citet{Poteet} studied the line of sight towards $\rm \zeta$ Ophiuchi, using infrared observations, and found that nearly all the Mg and Si atoms reside in amorphous silicate grains, while a substantial amount of iron resides in other compounds. Hydrodynamic simulations of \citet{Zhukovska2018} suggest that a large fraction of iron ($\rm 70\%$) is locked as inclusions in silicate grains, where it is protected from sputtering by SN shocks. The remaining depleted iron could reside in a population of metallic iron nanoparticles with sizes in the range of 1–10 nm. Other forms of solid-phase iron include metallic inclusions in glass with embedded metal and sulphides of interstellar origin (e.g \citealt{Keller}). 

Oxygen is one of the most abundant elements in the Galaxy. However, the total oxygen budget in the ISM is largely uncertain (\citealt{Jenkins2009}). While oxygen is expected to be mildly depleted into dust, a significant fraction of it is missing from the gas phase at a level that cannot be fully explained yet. The combined contributions of CO, ices and silicate/oxide dust cannot fully account for the missing oxygen in the dense regions of the ISM, and in particular at the interface between diffuse and dense ISM (\citealt{Whittet2010}, \citealt{Poteet}). A possible solution to the oxygen budget problem of the ISM has been proposed by \citet{Jones2019}. The authors conclude that a significant fraction of oxygen in the dense ISM could exist in organic carbonate solids ($\rm [^{C-O}_{C-O}C=O$). This material maximises the O/C atom ratio (3/1) so that oxygen could be efficiently depleted, without over-depleting carbon, the most likely elemental depleting partner of oxygen.

In this paper we study simultaneously the O K and Fe L edges using the most up-to-date dust laboratory measurements. This is the first time that these two elements are studied together with a new accurate dust extinction model, considering also the effect of the dust scattering (see paper I). We studied the iron and oxygen abundances and depletions\footnote{In this work, we define depletion of an element as the ratio of the dust abundance to the total abundance of a given element (both gas and dust)} in five sight-lines along the Galactic plane, with column densities of $\rm <5\times10^{21}\ cm^{-2}$. This paper is organized as follows. In Section \ref{sample} we present the sample of X-ray binaries and in Section \ref{data} we show the X-ray data used for this work. The methodology used for the spectral fitting of the ISM is explained in Section \ref{fitting}. Finally, in Section \ref{discussion} we discuss our results, and give our conclusions in Section \ref{conclusions}. 

\section{X-ray sources}
\label{sample}

To study the O K and Fe L edges simultaneously, we adopt five low-mass X-ray binaries (LMXBs) as background sources, suitable to model the ISM absorption. LMXBs are excellent candidates due to their high flux and the absence of emission features that could complicate the analysis. The depth and visibility of the X-ray photoabsorption edges is highly dependent on the ISM column density in the foreground of the X-ray binary. We therefore select sources with a column density between $\rm 1-5 \ \times 10^{21} cm^{-2}$ in order to guarantee a large optical depth for both the O K and Fe L edges. The flux limit has been chosen to be larger than $\rm 1 \times 10^{-9} erg^{-1}\ s^{-1} \ cm^{-2}$ in the 0.1-2 keV band to ensure a high signal-to-noise spectrum.

\begin{figure} [ht]
\includegraphics[width=0.5\textwidth]{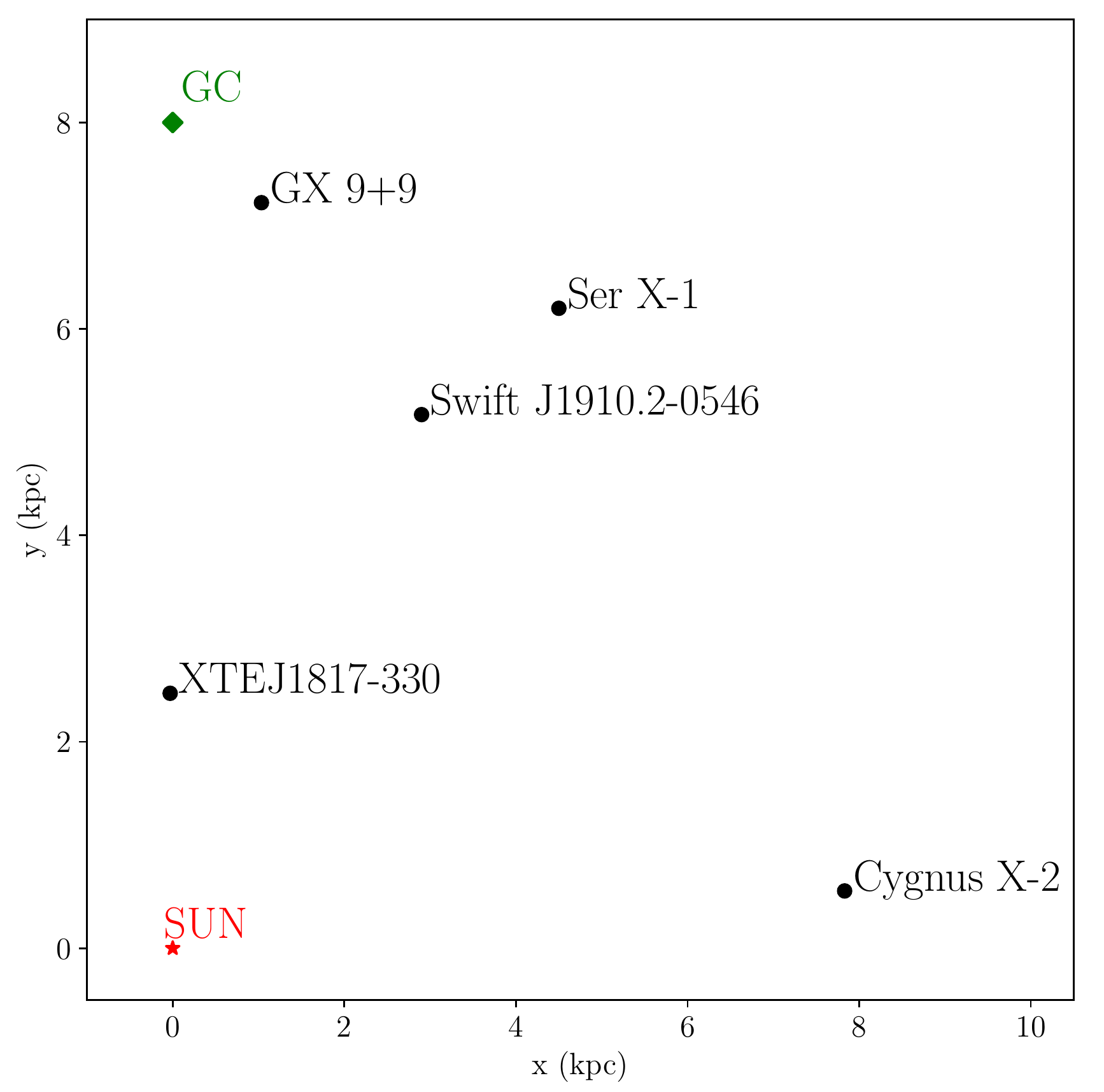}
\label{fig:distances}
\caption{Projection of the X-ray sources on the Galactic plane. The distance between the sun and the Galactic center is assumed to be 8 kpc.}
\end{figure}

\begin{figure} [ht]
\includegraphics[width=0.5\textwidth]{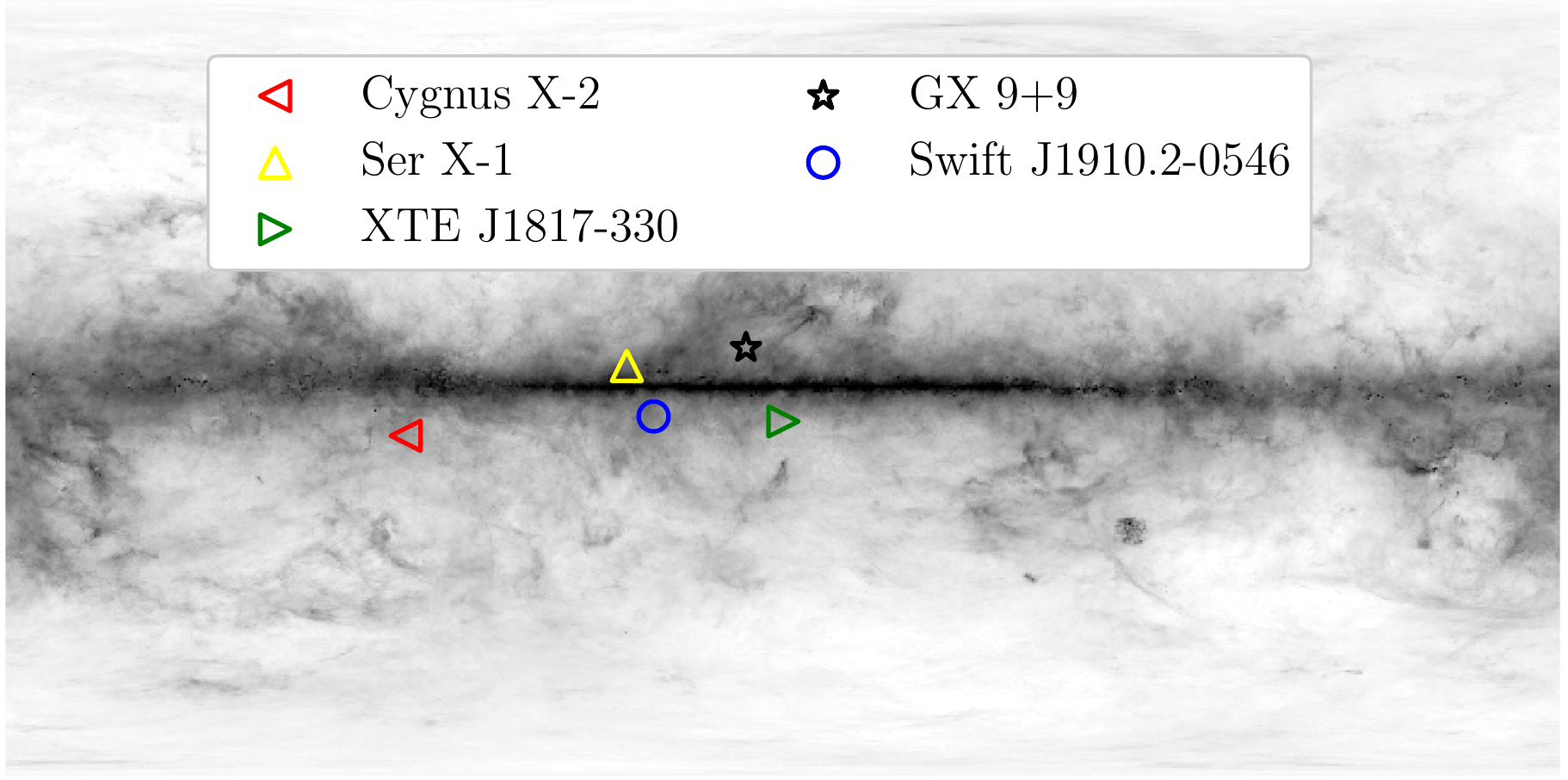}
\label{fig:galaxy}
\caption{The X-ray sources plotted on a 100 $\rm \mu m$ full-sky map, adopted from \citet{Schlegel}.}
\end{figure}

Another requirement is the availability of data from both \xmm and \chandra public archives. The selected sample of sources as well as the obs. ID and exposure times are listed in Tables \ref{xmmdata} \& \ref{chandradata}. In Table \ref{xmmdata} we also show the distance to the source, the column density ($\NH$) taken from the \citet{HI4PI}, along with the reddening values E(B-V), when available, and the calculated column density ($\NH^{E(B-V)}$) derived using the conversion $\NH/E(B-V)= \rm 5.8\times10^{21}H \ cm^{-2} \ mag^{-1}$ \citep{Drainebook}. In Fig. \ref{fig:distances} we present the projection of the X-ray sources on the Galactic plane. In Figure \ref{fig:galaxy} we plot the sources on an all-sky map of the galactic dust emission at a wavelength of 100 $\mu m$ (\citealt{Schlegel}). 

\begin{table*}
\caption{\xmm RGS observation log}  
\label{xmmdata}    
\renewcommand{\arraystretch}{1}
\begin{center}
 \small\addtolength{\tabcolsep}{+1.2pt}
\scalebox{0.98}{
\begin{tabular}{c c c c c c c c }     
\hline\hline            
Source                       &      obs. ID &   exposure time (ks)  &     distance (kpc)         &  Galactic coordinates (l,b)   &       \NH            & $E(B-V)$               & $\NH^{E(B-V)}$        \\
\hline   
\hline     
\multirow{2}{*}{Cygnus X--2} &  0303280101 &      32            & \multirow{2}{*}{8-11$^{(a)}$}   & \multirow{2}{*}{(87.3,-11.3)}  & \multirow{2}{*}{1.9}  & \multirow{2}{*}{$0.4\pm0.07^{(1)}$}& \multirow{2}{*}{$2.3\pm0.4$}            \\
                             &   0561180501&      24             &                                    &                                &                                \\
\hline                
\multirow{3}{*}{GX 9+9}      & 0694860301   &     36              &\multirow{3}{*}{ 5-7$^{(b)}$} &  \multirow{3}{*}{(8.5,9)}   &  \multirow{3}{*}{1.9} &  \multirow{3}{*}{0.1-0.4$^{(2)}$} &\multirow{3}{*}{0.6-2.3}              \\
                             &   0090340101 &     11              &                                   &                                &                                \\
                             &   0090340601 &     22              &                                   &                                &                                 \\
\hline                
\multirow{3}{*}{Ser X--1}    &   0084020401 &    22  & \multirow{3}{*}{$7.7\pm0.9^{(c)}$} &  \multirow{3}{*}{(36.1,4.8)} &  \multirow{3}{*}{4.4} &  \multirow{3}{*}{0.4-1.5$^{(3)}$}& \multirow{3}{*}{2.3-8.7}         \\
                             &   0084020501 &    22               &                                   &                                &                               \\
                             &   0084020601 &    22               &                                   &                                &                               \\
\hline                
\multirow{1}{*}{SWIFT J1910.2--0546} &   0691271401 &   28        &   6$^{(d)} $                       &   (29.9,-6.8)                  &   2.24          &   - & -              \\
\hline   
\multirow{1}{*}{XTE J1817--330}   &   0311590501 &     20         &   1-4$^{(e)}$             &   (359.8, -8)                  &    1.5          & 0.1-0.5$^{(4)}$ & 0.6-3               \\
\hline
\end{tabular}}
\end{center}
\tablefoot{The symbol $\NH$ represents the hydrogen column density in $\rm 10^{21} \ cm^{-2}$ from \citet{HI4PI}, and $\NH^{E(B-V)}$ is the column density in $\rm 10^{21} \ cm^{-2}$ estimated from the reddening values $E(B-V)$ taken from: $^{(1)}$\citet{McClintock}, $^{(2)}$\citet{Shahbaz}, $^{(3)}$\citet{Thorstensen}, $^{(4)}$\citet{Torres}. Distances are from: $^{(a)}$\citealt{Cowley} \& \citealt{Smale}, $^{(b)}$\citet{Christian} \& \citet{Vilhu2007}, $^{(c)}$\citet{Galloway}, $^{(d)}$\citet{Reis}, $^{(e)}$\citet{Sala}. To convert from $E(B-V)$ to \NH we used the conversion $\NH/E(B-V)= \rm 5.8\times10^{21}H \ cm^{-2} \ mag^{-1}$ \citep{Drainebook}.}
\end{table*}


\begin{table*}
\caption{\chandra observation log}  
\label{chandradata}    
\renewcommand{\arraystretch}{1.2}
\begin{center}
 \small\addtolength{\tabcolsep}{+2pt}
\scalebox{1}{%
\begin{tabular}{c c c c  }     
\hline\hline            
Source                       &     obs. ID  &      instrument/mode        &      exposure time (ks)     \\
\hline   
\hline     
\multirow{2}{*}{Cygnus X--2} &  8170        &    HETGS/CC                 &    65                             \\

                             &  8599        &    HETGS/CC                 &     60                             \\                         

\hline                
\multirow{2}{*}{GX 9+9}      & 11072      & HETGS/TE                     &    95                                          \\
                             &   703      & HETGS/TE                     &    20                                         \\
\hline                
\multirow{1}{*}{Ser X--1}    &  700       &    HETGS/TE                  &    77                                  \\

\hline                
\multirow{1}{*}{SWIFT J1910.2--0546} &   14634 &  HETGS/CC                   &    30                                       \\

\hline                
\multirow{1}{*}{XTE J1817--330}   &   6617 &   HETGS/CC                     &    47                                           \\

\hline                
\end{tabular}}
\end{center}
\end{table*}


\section{Data processing and reduction}
\label{data}
In order to best study the narrow absorption features of gas and dust present in the soft X-ray band (< 1 keV) we combine the capabilities of currently available high-resolution X-ray spectrometers on board \xmm and \chandra satellites. To study the oxygen K-edge region we use the Reflection Grating Spectrometers (RGS, \citealt{denHerder}) of \xmm which has a resolving power of $R=\frac{\lambda}{\Delta \lambda } \gtrsim 400$ and an effective area of approximately $\rm 45 \ cm^{2}$ in this spectral region.

\chandra carries a high spectral resolution instrument, namely High Energy Transmission Grating \cite[HETGS,][]{hetgs}. The HETGS consists of two sets of gratings, the Medium Energy Grating (MEG) and the High Energy Gratings (HEG). For the iron L-edges we use \chandra data, due to the higher spectral resolution of HETGS/MEG around the Fe L-edges. MEG has an effective area of approximately $7 \ {\rm cm}^{2}$ and a resolving power of $R=\frac{\lambda}{\Delta \lambda } \gtrsim 760$ around the iron L-edges which makes it the most suitable instrument to study the XAFS with archival data in this spectral region. A similar approach, using combined information from \xmm and \chandra data has been already presented in \citet{Costantini2012}. All the observation IDs (obsid) used in this paper are listed in Tables \ref{xmmdata} and \ref{chandradata} .

\subsection{\xmm}
\label{xmm}

We obtain the RGS data from the XMM-Newton public archive\footnote{\url{http://nxsa.esac.esa.int/nxsa-web/}}. The data have been reduced using the Science Analysis Software, $\textit{SAS}$ (ver.18). We first run the $rgsproc$ command to create the event lists. Then, we filter the RGS event lists for flaring particle background using the default value of 0.2 counts/sec threshold. The bad pixels are excluded using keepcool=no in the SAS task $rgsproc$. In bright sources, some areas of the grating data may be affected by pile-up, which can change the shape of the spectrum below 19 \AA. To avoid the effect of pile-up, we ignore the region of the spectra shorter than 19 \AA. In this analysis, we are cautious with combining different observations of the same source in order to avoid artifacts in the spectrum created from the merging. However, when the spectral shape and flux of certain observations does not vary, or the data have been acquired in the same epoch and can be superimposed, we combine them using the SAS command $\textit{rgscombine}$. This allows us to obtain a single spectrum with higher signal-to-noise ratio. 
\subsection{\chandra}
\label{chandra}

The \chandra observations used in this work were downloaded from the Transmission Grating Catalogue\footnote{\url{http://tgcat.mit.edu/tgSearch.php?t=N}} (TGCat, \citealt{tgcat}). For each observation, we combine the positive and negative spectral orders using the X-ray data analysis software, CIAO (version 4.11, \citealt{ciao}). Depending on the availability of the data, we use observations taken in either timing mode (TE) or continuous clocking (CC) mode. In bright sources, to avoid pile-up we ignore the region of the spectra below 10 \AA. Similarly with the \xmm observations, if the spectral shape of a certain source is steady, and the spectra do not present any variations, we combine the different observations using the CIAO tool $\combinespectra$. 

\section{Spectral fitting of the ISM}
\label{fitting}
\subsection{The two-edge fit}

In this work, we fit simultaneously the O K- and Fe L-edges. For our modelling we use the software SPEctral X-ray and UV modelling and analysis, \spex, version 3.06.01\footnote{\url{http://doi.org/10.5281/zenodo.2419563}} (\citealt{Kaastra1996, Kaastra2018}). We fit the \chandra data in the range of 10-19 $\AA$ and the \xmm spectrum between 19-35 $\AA$. This wavelength range covers an area which is sufficiently broad around the edges and excludes areas that can complicate our fit and hamper the XAFS modelling. 

The \xmm and \chandra data of each source were obtained at different epochs and therefore the continuum shape might be different due to variability intrinsic to the source. To take this variability into account, we use the \sectors option in \spex. With this option, each dataset is allocated to a specific sector. In this way, the continuum parameters (which are variable over time) for each dataset is treated independently and allowed to vary freely while the ISM model is jointly fitted to all the data. Because we use a narrow energy band for each observation, we do not require fitting the broadband continuum with any physical models. Hence, we use a phenomenological power law model (\pow in \spex) and a black body component (\bb) to describe the continuum in this energy range.

We apply a binning on our data of a factor of 2, which improves the signal to noise while the data are still oversampling the spectral resolution of the instrument and we are not loosing accuracy. We further adopt $C$-statistics ($C_{\rm stat}$) to evaluate the goodness of our fit (\citealt{Cash1979}, \citealt{Kaastra2017}). All uncertainties are provided at 1$\rm \sigma$ significance. Also, in our analysis we use proto-Solar abundance units from \citet{Lodders2009}. A step-by-step description of our spectral fitting procedure is presented in Section 4.2.

\subsection{Fitting procedure}
\label{fitiing2}

The fitting procedure described in this section is applied to all the sources presented in Figure \ref{fig:distances}. \\

{\bf The multi-phase gas modelling}\\

To characterize the neutral galactic absorption, we adopt the \hot model of \spex (\citealt{dePlaa}). For a given temperature and set of abundances, this model calculates the ionisation balance and then determines all the ionic column densities by scaling to the prescribed total hydrogen column density. At low temperatures ($\sim$ 0.001 eV $\sim$  10 Kelvin) it mimics the neutral gas. The free parameters here are the hydrogen column density (\NH) and the temperature ($kT$, where $k$ is the boltzmann constant) as well as the relative abundance of oxygen, iron, silicon and magnesium.

Two additional \hot models are used to take into account the weakly and mildly ionised gas. In this way, we probe the different ISM phases along the line of sight of each source. The column densities and temperatures of the \hot components derived from our spectral fitting are listed in Table \ref{hotparameters}.

We further take into account the absorption from a hot gas (e.g. \ovii, \oviii) using the \slab model in \spex. This model calculates absorption by a slab of optically-thin gas, where the column density of ions are fitted individually and are independent of each other. The nature of these absorption lines is still debated. They have been attributed to either warm absorbers intrinsic to the binary system, or from a hot phase of the ISM. The investigation of the origin of these lines is outside the scope of this paper, and therefore we do not provide a physical interpretation for them.\\

%
\begin{table*}
\caption{Best fit parameters of the ISM gas-phase components.}  
\label{hotparameters}    
\renewcommand{\arraystretch}{1}
\begin{center}
 \small\addtolength{\tabcolsep}{+2pt}
\scalebox{1.1}{%
\begin{tabular}{c c c c c}     
\hline\hline            
Source                        &                & \hot \# 1 &   \hot \# 2  &     \hot \# 3   \\
\hline   
\hline   
 \multirow{2}{*}{Cygnus X--2} &  \NH           &$1.7\pm0.3$  &$0.04\pm0.01$ & $0.027\pm0.007 $    \\
 							  &   $kT$         &0.001 $(f)$          &$2.6\pm0.6$     & $14.7\pm0.1 $            \\
\hline                
\multirow{2}{*}{GX 9+9}      &\NH              & $2.1\pm0.1 $ & $0.10\pm0.02$ &   $0.027\pm0.005$ \\
 							  &   $kT$         & 0.001 $(f)$        & $2.6\pm0.8 $  &   $210\pm23$        \\

\hline                
\multirow{2}{*}{Ser X--1}    &  \NH            & $5.5\pm0.1$        &$0.4\pm0.1$       & $0.11\pm0.05$  \\
 							  &   $kT$         & 0.001 $(f)$             & $2.2\pm0.8$    & $14\pm2$    \\

\hline               
\multirow{2}{*}{SWIFT J1910.2--0546} & \NH     &$3.20\pm0.04$ &  $0.03\pm0.01$  &  $0.022\pm0.005$            \\
 							  &   $kT$         & 0.001 $(f)$       &  $13\pm1$        &  $145\pm10$               \\

\hline   
\multirow{2}{*}{XTE J1817--330}  & \NH         & $1.50\pm0.01$  & $0.09\pm0.02$     &  $0.03\pm0.01$     \\
 							  &   $kT$         & 0.001 $(f)$               & $3.0\pm0.7$       & $10\pm2$             \\

\hline
                
\end{tabular}}
\end{center}
\tablefoot{Best fit parameters of the 3 \hot components. The symbol \NH represents the hydrogen column density in $\rm 10^{21} \ cm^{-2}$ and $kT$ represents the temperature in eV. The temperature of the neutral gas component (hot \# 1) is frozen to the minimum value in order to produce neutral species only. This is noted with the symbol $f$.}   
\end{table*}

{\bf The dust modelling}\\

In addition to the gas components, we model the dust absorption in the O K and Fe L edges with a custom version of the \amol model in \spex. This model calculates the transmission of a dust component, and the free parameter is the dust column density. We use the recently implemented dust extinction cross sections for the oxygen K-edge presented in Paper I. For the iron L-edges we use new dust extinction cross sections computed using the same laboratory measurements and methodology (Costantini et al. in prep). The laboratory experiment and post-processing are explained in detail in Paper I \cite[also see][]{Zeegers2017, Rogantini2018}. In summary, the total extinction for oxygen has been calculated using Mie theory (\citealt{Mie}), and in particular the python library $miepython$\footnote{\url{https://miepython.readthedocs.io/en/latest/}}. For the iron models, the total extinction cross section has been calculated using Anomalous Diffraction Theory \cite[ADT,][]{van_de_Hulst}. For both oxygen and iron, a Mathis-Rumpl-Nordsieck dust size distribution has been assumed \cite[MRN,][]{mathis}, which follows a power-law distribution, $dn/da\propto a^{-3.5}$, where $a$ is the grain size with a minimum cut-off of $\rm 0.005 \mu m$, and a maximum of $\rm0.25 \mu m$. \\

\textit{\large The minerals.}\ \ \  In Table \ref{tab:samples}, we summarize the different types of dust mineralogy used in this work by specifying their chemical formula and lattice structure (i.e. crystallinity). Our dust samples include silicates, such as olivines and pyroxenes, with varying Mg:Fe ratio. We also include oxides, iron sulfides and metallic iron. The latter has been adopted from the literature (\citealt{Kortright}, \citealt{Lee2010}) and has been shifted according to the energy calibration value supplied by \citet{Fink}. Similarly with all the compounds, the scattering component of extinction has also been included here, using the MRN dust size distribution as mentioned above.\\
%
\begin{table}
\caption{List of dust samples.}
\begin{tabular}{cccc}
  \hline
  \hline
 \# & Compound & Chemical Formula & Form\\
  \hline
  1 &  Olivine & $ \rm MgFeSiO_{4}$ &  amorphous \\
  2 &  Olivine & $ \rm Mg_{1.56}Fe_{0.4}Si_{0.91}O_{4}$ &  crystalline \\
 3 &  Pyroxene & $ \rm Mg_{0.6}Fe_{0.4}SiO_{3}$ &  amorphous \\
 4 &  Pyroxene &  $ \rm Mg_{0.6}Fe_{0.4}SiO_{3}$ &  crystalline \\
 5 & Enstatite & $ \rm MgSiO_{3}$ &  crystalline \\
  6 & Enstatite & $ \rm MgSiO_{3}$&  amorphous \\
 7 &  Fayalite & $ \rm Fe_{2}SiO_{4}$ &  crystalline \\
 8 &  Forsterite & $ \rm Mg_{2}SiO_{4}$ & crystalline \\
 9 & Pyroxene & $ \rm Mg_{0.75}Fe_{0.25}SiO_{3}$ &  amorphous \\
 10 & Magnetite & $ \rm Fe_{3}O_{4}$ &  crystalline \\
 11 &  Troilite & $ \rm FeS$ &  crystalline\\
 12 &  Pyrit Peru & $ \rm FeS_{2}$ &  crystalline\\
 13 &  Metallic iron & $ \rm Fe$ &  -         \\
\hline
\hline
\end{tabular}
\tablefoot{The samples 2,5,11 are natural and 8,10,12 are commercial products. Samples 1,3,4,6,7,9 are instead synthesized in the laboratories at the Astrophysikalisches Institut, Universitats-Stenwarte (AIU), and Osaka University. We adopted the metallic iron presented by \citet{Kortright} and \citet{Lee2010}, with a shift in energy according to the work of \citet{Fink}.}
\label{tab:samples}
\end{table}

\textit{\large Fitting method.}\ \ \   The \amol model can fit up to four different dust compounds simultaneously at a given fitting run. We test all the possible combinations of four dust species among the 13 samples assuming that the interstellar dust mixture can be described with at most 4 components. The number of different combinations is given by the equation:

\begin{align*}
\rm C_{e,c}=\frac{e!}{c!(e-c)!}
\end{align*}

\noindent where $e$ is the number of the available edge profiles and $c$ the combination class. This gives us 715 dust mixtures to choose from when fitting the spectra for each source. 

\begin{figure*} 
\begin{center}
\includegraphics[width=1\textwidth]{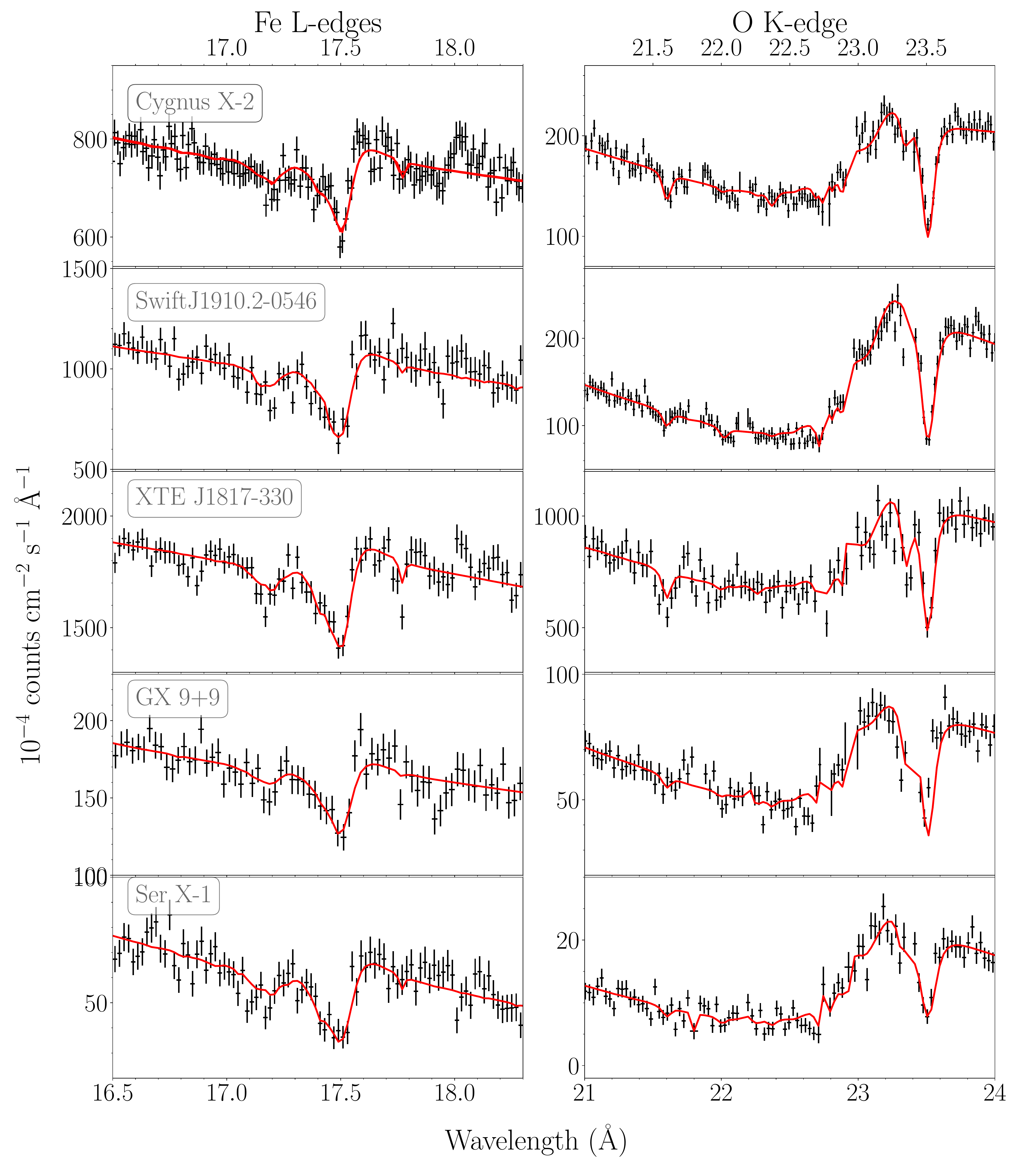}
\label{fig:bestfit2}
\caption{Best fit in the O K and Fe L-edges. \xmm data has been used to fit the O K-edge spectral region, while for the Fe L-edges we use spectroscopic data from \chandra. The spectral fitting has been done by taking into account both gas (neutral+ionised) and dust absorption. }
\end{center}
\end{figure*}

\begin{figure*} 
\begin{center}
\includegraphics[width=1\textwidth]{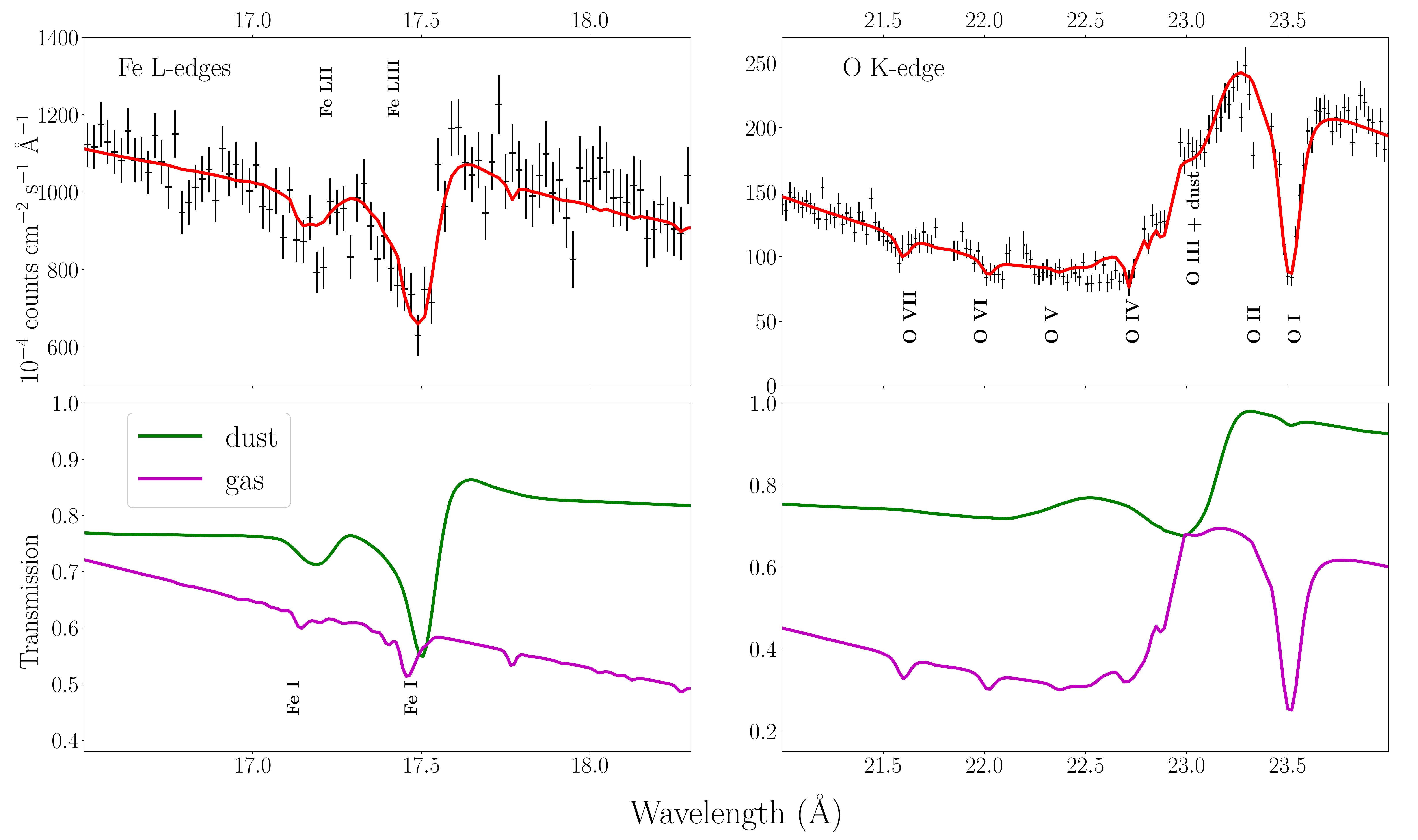}
\label{fig:transmission}
\caption{Best fit in the O K and Fe L-edges for SWIFT J1910.2--0546 and the relative transmission for the gas and dust components. The transmission of the gas has been multiplied by a factor of 2.5 and 4.5, for iron and oxygen respectively for display purposes. }
\end{center}
\end{figure*}

In Figure \ref{fig:bestfit2} we present, for all the sources, the best fit in the O K- and Fe L-edges. In Table \ref{tab:bestfitdust} we list the best fit dust mixtures with the column densities of each dust compound contributing to the absorption. In the same Table we also present the calculated dust-to-gas mass ratio ($D$) for every line of sight. The dust mass has been calculated accounting the mass of the elements that make up the best fit minerals, while we assume that the gas mass is dominated by hydrogen and helium. In Figure \ref{fig:transmission} we present the best fit for a selected source, SWIFT J1910.2--0546, in detail. The bottom panel shows the transmission of gas and dust components used in the fit. The Fe L-edge shape is dominated by the dust absorption, while the fit to the O K edge is dominated by gas features. However, the dust absorption in the O K edge is not negligible, it covers a broad region around the edge an improves the fit around 23.3 \AA, where the scattering component of extinction is present (see Paper I).\\

%

\textit{\large Model selection.}\ \ \    Following the methodology used in Paper I and in \citet{Rogantini2020}, we evaluate the best fit among the various dust mixtures using the \textit{Akaike Information Criterion} (AIC, \citealt{Akaike}). AIC provides a robust and fast way to select the models that are statistically similar to the best fit. Traditional likelihood-based methods for model selection, provide a mechanism for estimating the unknown parameters of a model, having a specified dimension and structure. \citet{Akaike} extends this paradigm by considering a framework in which the model dimension is also unknown (\citealt{Feigelson}). AIC treats all the models symmetrically, not requiring an assumption that one of the candidate models is the correct model. It can be therefore used for both nested and non-nested models. In our case, the models are non-nested and with the same number of free parameters. In practice, this method allows a quick comparison of the candidate's models by comparing the C-statistic value of every fit with the best one. The AIC value is given by:

\begin{align*}
\rm AIC=2\textit{k}-2ln(\La_{max})
\end{align*}

\noindent where $k$ is the number of fitted parameters of the model and $\La_{max}$ is the maximum likelihood value. The C-statistics value and the maximum likelihood are related as $\rm Cstat=-2ln(\La)$ (\citealt{Cash1979}). We are not directly interested in the AIC value; we calculate the AIC difference ($\rm \Delta AIC$) over all candidate models with respect to the model that has the lowest AIC value. Models with $\rm \Delta AIC<4$ are considered comparable and can fit the spectrum equally well, while models with $\rm \Delta AIC>10$ can be ruled out (\citealt{Burham}). 

\begin{figure*} 
\begin{center}
\includegraphics[width=0.90\textwidth]{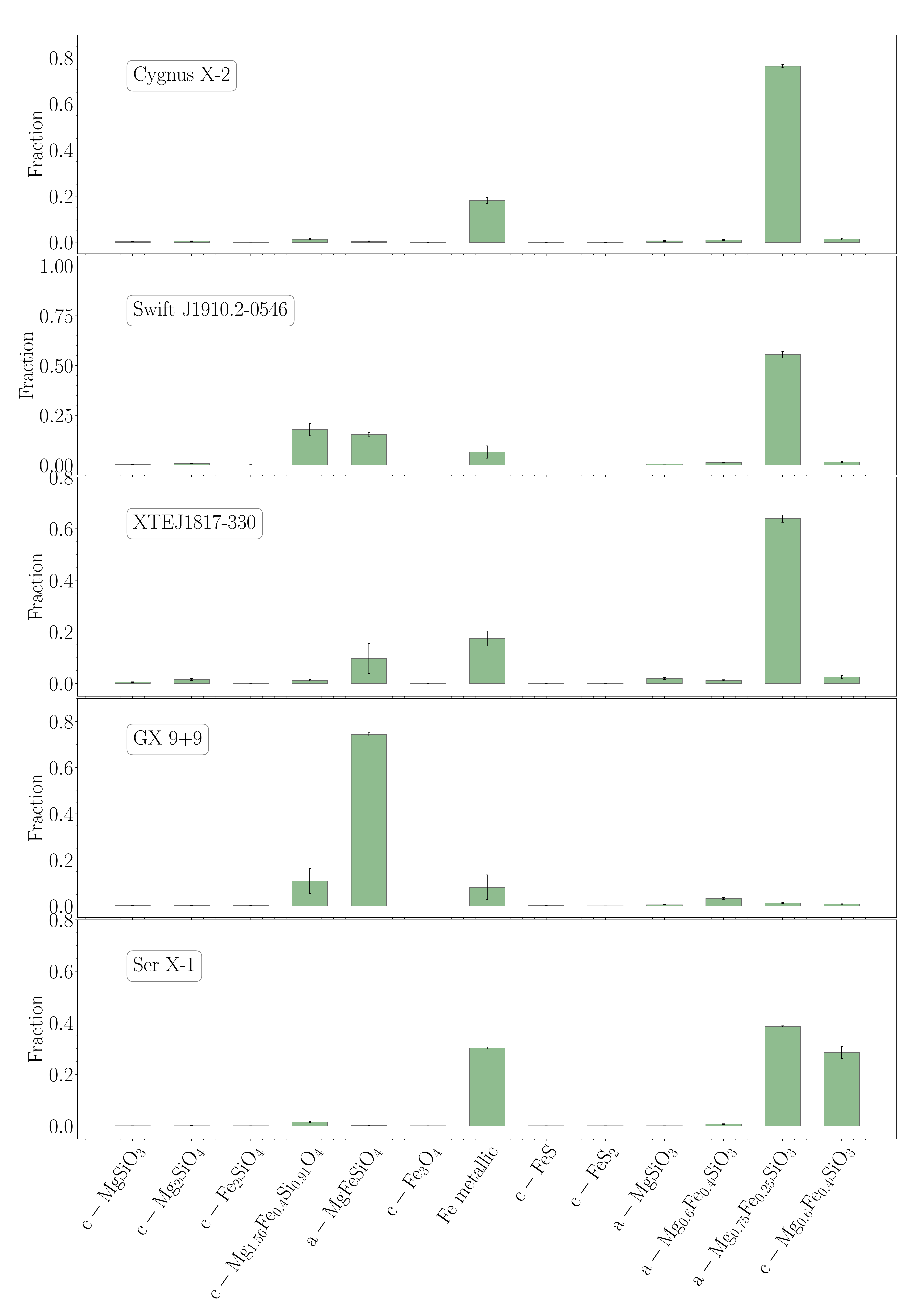}
\label{fig:barplot1}
\caption{Relative fraction of column density for each dust compound. The fraction has been calculated considering models with $\rm \Delta AIC <4$. The symbol $a$ refers to amorphous compounds and $c$ to crystalline. The errors on the bar charts is an indication of the minimum and maximum percentage of a given compound, within the $\rm \Delta AIC<4$ selected models. The values of the least significant compounds is an upper limit. }
\end{center}
\end{figure*}

In Figure \ref{fig:barplot1} we present the relative fraction of each dust compound using the AIC criteria. In particular, in the bar chart we show the results for the models with $\rm \Delta AIC<4$ including all the fits with similar significance compared to the best fit (about 50 on average). 
The errors on the bar charts is an indication of the minimum and maximum percentage of a given compound. The values of the least significant compounds is an upper limit. 

%

\begin{table*}
\caption{Dust column densities for each chemical compound in $\rm 10^{16} \ cm^{-2}$ which corresponds to the best fit of each source.}
\label{tab:bestfitdust}
\begin{center}
 \scalebox{1}{%
\begin{tabular}{clccccc}
  \hline
  \hline
 Compound \# &  Dust compound								     & Cygnus X--2  & GX 9+9    & Ser X--1 & SWIFT J1910.2--0546 & XTE J1817--330  \\
  \hline
 1 			&  $ \rm \textit{a}-MgFeSiO_{4}$	 				 &-              &$4.5\pm0.8$&-           &	  			       & -          \\
 2 			&  $ \rm \textit{c}-Mg_{1.56}Fe_{0.4}Si_{0.91}O_{4}$ &-			  	&<1.2      &	-		  &	$5.3\pm3.4$		   &-                  \\
 3 			&   $ \rm \textit{a}-Mg_{0.6}Fe_{0.4}SiO_{3}$        &<1.8			 &	-	  &	-		      &	-				   &<1                  \\
 4 			&   $ \rm \textit{c}-Mg_{0.6}Fe_{0.4}SiO_{3}$        &-			  	 &	-	  &	-		      &		               & <3                \\
 5 			&   $ \rm \textit{c}-MgSiO_{3}$                      &-				 &	-	  &			      &	-				   & -        \\
 6 			&  $ \rm \textit{a}-MgSiO_{3}$                       &-               &<0.4    &	-             &	-			       & -                 \\
 7 			&   $ \rm \textit{c}-Fe_{2}SiO_{4}$                  &-			    &	- &-              &	-			       &  -                \\
 8 			&   $ \rm \textit{c}-Mg_{2}SiO_{4}$                  &-			   &	-	  &	-		      &	-				   &  -      \\
 9 			&   $ \rm \textit{a}-Mg_{0.75}Fe_{0.25}SiO_{3}$      &$8\pm1$     	&	-	  & $17.1\pm0.2$  &	 $9\pm4$           &  $8\pm1$              \\
 10 		&  $ \rm \textit{c}-Fe_{3}O_{4}$                     &-			  	  &-    &	<0.9		  &	-				   &    -              \\
 11 		&  $ \rm \textit{c}-FeS$                             &-            	  &	-	  & -             &	-			       &     -             \\
 12 		&  $ \rm \textit{c}-FeS_{2}$                         &<1            	&	-	  &<0.5             &	<0.2			       &      -    \\
 13 		&  $ \rm Fe \ metal $                       		 &$1.8\pm0.9$	   &	<1.3  &	$8\pm2$		  &	$3\pm1$			   &   $2.3\pm0.7$     \\
 \hline
  $C_{\rm stat}$/$d.o.f.$            &                           &1311/1056    &2098/1667      &1005/787       & 2122/1837         &  1188/1002         \\
  \hline
  $D$                     &                        &0.005         & 0.004     &0.003   & 0.004         & 0.007       \\

\hline
\hline
\end{tabular}}
\end{center}
\tablefoot{The symbol $a$ refers to amorphous compounds and $c$ to crystalline.In the last rows we present the $C_{\rm stat}$ of the best fit in every source compared to the degrees of freedom ($d.o.f.$) and the dust-to-gas mass ratio ($D$).}
\end{table*}
%
%

%
%
\section{Discussion}
\label{discussion}

\subsection{Silicate mineralogy}
\label{mine}

Our analysis of the 5 X-ray binaries shows that Mg-rich amorphous pyroxene ($\rm Mg_{0.75}Fe_{0.25}SiO_{3}$) is the dominant compound in most sight-lines. In Figure \ref{fig:barplot1} we present, for every source, the relative fraction of column density of each contributing to the dust content chemical compound. For this plot, we consider models with $\rm \Delta AIC <4$. In particular, we find that Mg-rich amorphous pyroxene accounts for approximately $\rm \sim 70\%$ of the dust column towards Cygnus X-2, Swift J1910.2-0546, XTE J1817-330 and Ser X-1.

The results on the silicate mixture of dust are broadly consistent with previous studies at the infrared wavelengths. \citet{Min2007} studied the spectral shape of the $\rm 10 \ \mu m $ interstellar extinction feature towards the Galactic Center. They found that interstellar silicates are magnesium rich and that the stoichiometry lies between pyroxene and olivine types. The authors suggest that the high magnesium content of amorphous silicates provides an explanation for the relatively high amount of magnesium found in cometary and circumstellar grains. Other infrared studies have also shown that the silicate feature can be modeled with a mixture of Mg-rich olivines and pyroxenes (\citealt{Molster2002}, \citealt{Chiar2006}).

In our study, we find a different dominant silicate mineralogy than what was previously discovered in \citet[][henceforth Z19]{Zeegers2019} and \citet[][henceforth R20]{Rogantini2020}, using the same dust laboratory samples. Our analysis shows that amorphous pyroxene ($\rm Mg_{0.75}Fe_{0.25}SiO_{3}$) is the dominant dust compound. Z19 and R20 studied the gas and dust absorption on the silicon and magnesium K-edges showing that amorphous olivine ($\rm MgFeSiO_{4}$) represents most of the dust in the dense environments of the ISM ($\rm \sim 10^{22} \ cm^{-2}$). Here, we probe more diffuse regions of the ISM ($\rm \sim 10^{21} cm^{-2}$) and different lines of sight and therefore variations in the results are possible.

In fact, \citet{Demyk2001} performed irradiation experiments exposing crystalline olivine and pyroxene materials to high energy ions. They found that under $\rm He^{+}$ irradiation the olivine loses Mg and O and therefore the silicate composition could evolve from olivine to pyroxene in the diffuse ISM. Our lines of sight probe regions with typical gas densities that are an order of magnitude lower than those of R20 and Z19. It is possible that the degree of processing of silicates along these different sight-lines differs. Our sight-lines could represent a population of dust that is losing oxygen and magnesium atoms to the gas phase, which could lead to a more pyroxene like stoichiometry. In other, higher density environments this process may be less efficient, or accretion of atoms to dust grains may become important.

GX 9+9 is an outlier. Amorphous olivine ($\rm MgFeSiO_{4}$) is the dominant compound in this line of sight, contrary to amorphous Mg-rich pyroxene ($\rm Mg_{0.75}Fe_{0.25}SiO_{3}$) found towards the other X-ray sources. It is possible that with GX 9+9 we are probing a different ISM environment. We examined the three-dimensional maps of interstellar dust reddening, which are based on Pan-STARRS 1 and 2MASS photometry, and Gaia parallaxes\footnote{\url{http://argonaut.skymaps.info/}} (\citealt{Green2019} and references therein). These maps trace the dust reddening both as a function of angular position on the sky and distance. Using these maps, we found that there is a steep jump in the line-of-sight reddening towards GX 9+9. This suggests that there is a dense cloud of dust in front GX 9+9. The profiles of reddening versus distance for the other targets in this study do not generally exhibit steep jumps. Instead, the amount of reddening increases more gradually with distance, suggesting that the dust along these sight lines is more diffuse in its distribution. We therefore conclude that the GX 9+9 sight line is different from the others, possibly containing a denser cool cloud, which might explain why amorphous olivine fits this spectrum better.

We find an upper limit on the silicate crystallinity of 15\%. Our results are consistent with R20 and Z19. The authors found that $0-30 \%$ of dust is in crystalline configuration, depending on the line of sight. Additionally, the shape of the $9.7 \mu \rm m$ and $18 \mu \rm m$ features in the IR suggests that the upper limit of crystallinity is constrained to 2.2\% (\citealt{Kemper2004}). If there is a discrepancy between the infrared and the X-ray band, it could be attributed to the nature of X-rays. X-rays are sensitive to a short range order of atoms and therefore we may observe crystallinity in partly glassy materials. Moreover, X-rays provide a probe of larger grains, while the infrared emission is dominated by the smaller dust sizes. It is therefore possible that X-rays are penetrating a portion of the grain that is shielded from cosmic rays, which destroy the crystalline structure.

\subsection{Abundances and depletions of O and Fe}
\label{abund}

We have calculated the abundances and depletions of oxygen and iron for the different lines of sight probed in this study, and we present the results in Table \ref{abundances}. The calculations are based on the parameters of the best fit model. In the first column, we list the sources used in this study. In the second ($N_{gas} $) and third column ($N_{dust}$), we list the gas and dust column density respectively. We note that the dust column density for O and Fe can be calculated from the values in Table \ref{tab:bestfitdust}. For example column densities of oxygen molecules in Table \ref{tab:bestfitdust} that include 3 oxygen atoms can be multiplied by 3 to get the total dust columns listed in Table \ref{abundances}. In the next two columns we show the abundances of the neutral gas and dust with respect to hydrogen. Column 6 shows the total abundance of gas and dust with respect to proto-Solar abundances of \citet{Lodders2009} and in column 7 we present the depletion of each element into dust, i.e., the fraction of oxygen and iron that has been locked up in dust grains with respect to the total. 

We find that about 10-20\% of the neutral oxygen is depleted into dust. In most lines of sight oxygen is slightly over-abundant, about 1.2-1.4 times the Solar value. Our results are consistent with previous studies of the O K and Fe L spectral regions. \citet{Costantini2012} studied the gas and dust absorption towards the galactic X-ray binary 4U1820-30, by simultaneously fitting the O K and Fe L-edges, with \xmm and \chandra observations, respectively. They found that oxygen is slightly over-abundant, by a factor of 1.23 times the Solar value, and 20\% of oxygen is depleted into dust. \citet{Pinto2013} studied the interstellar medium composition towards nine galactic LMXBs. They authors have studied the O K and Fe L-edges simultaneously using \xmm observations. They found that a significant amount of oxygen and iron is contained into solids, 15-25\% and 65-90\% respectively.

More than 90\% of iron in our study is depleted into dust grains. Within the errors, iron is found to be consistent with the Solar value, or slightly under-abundant. The under-abundance of iron might be attributed to the size distribution of dust grains studied here. In this study we are using an MRN dust size distribution ($\rm 0.005-0.25 \mu m$, \citealt{mathis}). However, some of the depleted iron could be included in larger grains with sizes above 0.25 $\rm \mu m$.

Self-shielding could contribute to the reduction of the total iron column available for photoelectric absorption (\citealt{Wilms}). In this case, strong absorption prevents X-rays from penetrating the inner portions of the dust grain, and a smaller fraction of the total metal column contributes to the absorption edge (\citealt{Corrales2016}). Self-shielding is included in our study, since we have taken into account the total extinction (scattering + absorption) cross section into our spectral modelling. However, this effect is expected to become noticeable for regions of the ISM that contain large grains, closer to the upper limit of the dust size distribution used in this study. It is therefore possible, that some of the depleted iron resides in large grain populations, greater than 0.25 $\rm \mu$m. This will be examined in a future study.

\begin{table*}
\caption{Oxygen and iron column densities, abundances and depletions.}  
\label{abundances}    
\renewcommand{\arraystretch}{1.1}
\small\addtolength{\tabcolsep}{+2pt}
\scalebox{1.1}{%
\begin{tabular}{c c c c c c c c}     
\hline\hline   
\multicolumn{7}{c}{\textbf{Oxygen}}                                                                   				                \\
\hline \hline          
\multirow{2}{*}{Source}                     &   $N_{gas} $                &  $N_{dust} $                &  $A_{gas} $        &	$A_{dust} $ 	     & $A/A_{\odot}$ & $\delta_{X}$  \\
                                            & $ \rm 10^{18} \ cm^{-2}$    &  $  \rm 10^{17} \ cm^{-2}$  &  $\rm 10^{-4}\ H^{-1}$ &  $\rm \rm 10^{-4}\ H^{-1}$         &               &               \\
\hline    
 \multirow{1}{*}{Cygnus X--2} 				&$1.1\pm0.1            $    &$2.5\pm0.3$			         & $6.4\pm0.6$          & $1.5\pm0.2 $    &$1.3\pm0.1$  &  $0.18\pm0.02$          \\
 \hline   
\multirow{1}{*}{XTE J1817--330} 			&$1.06\pm0.06$                 &$2.7\pm0.4$                 &$7.1\pm0.4$            &$1.6\pm0.3$		 &$1.4 \pm0.1$         & $0.18\pm0.03$       \\
\hline                
\multirow{1}{*}{Ser X--1}   				&$3.4\pm0.1	$                    &$5\pm2$    			     & $6.2\pm0.2$           &$1.0\pm0.4$	   &$1.20\pm0.07$     &$ 0.13\pm0.05$         \\
\hline    
\multirow{1}{*}{SWIFT J1910.2--0546} 		&$2.3\pm0.1$                &$4.8\pm1.7$                  & $7.30\pm0.03$           &$1.5\pm0.6	 $ & $1.4\pm0.1 $    & $0.17\pm0.06$         \\
\hline               
\multirow{1}{*}{GX 9+9}     				& $1.6\pm0.1	$               &$2.2\pm0.4$          	     & $7.9\pm0.5$          &$1.1\pm0.3$ 	   & $1.45\pm0.12$      &$0.12\pm0.02$            \\
\hline \hline    
\multicolumn{7}{c}{\textbf{Iron}}                                                                   				                \\
\hline\hline           
\multirow{2}{*}{Source}                     &   $N_{gas} $                &  $N_{dust}           $      &  $A_{gas} $         &	$A_{dust} $ 	      & $A/A_{\odot}$&  $\delta_{X}$         \\
                                            & $ \rm 10^{18} \ cm^{-2}$    &  $  \rm 10^{17} \ cm^{-2}$  &  $\rm 10^{-6}\ H^{-1}$       &  $\rm 10^{-5}\ H^{-1}$  &               &                   \\
 \hline\hline
 \multirow{1}{*}{Cygnus X--2} 				&$0.004\pm0.001$            &$0.4\pm0.1 $                &$2.6\pm0.7$            &$2.4\pm0.5$     	  & $0.79\pm0.16$          &$0.98\pm0.02$      \\
\hline   			
\multirow{1}{*}{XTE J1817--330} 			&$0.007\pm0.002$              &$0.4\pm0.1$                 &$2\pm1$      &$2.9\pm0.5$      &$1.02\pm0.16$                  &$0.93\pm0.16$              \\
\hline                
\multirow{1}{*}{Ser X--1}   				&$0.003\pm0.002$             &$1.20\pm0.05$                &$0.4\pm0.7$          &$2\pm1$      			& $0.8\pm0.3$            &$0.98\pm0.03$               \\
\hline                
\multirow{1}{*}{SWIFT J1910.2--0546} 		&$0.010\pm0.004$            &$0.7\pm0.3$                  & $3.3\pm1.3$     &$2.3\pm0.1$     	   & $0.8\pm0.3 $            & $0.87\pm0.08$        \\
\hline                
\multirow{1}{*}{GX 9+9}     				&$<0.005$                  &$0.5\pm0.1$          		  &<2.5               &$2.5\pm0.6$     		&$0.8\pm0.1$          & $>0.98$       \\
\hline        
\end{tabular}}    
\tablefoot{$N_{gas}$ and $N_{dust}$ correspond to the total column density of gas and dust respectively for each element (O and Fe). $A_{gas}$ and $A_{dust}$ indicate the abundances of oxygen and iron in gas and dust (with respect to hydrogen). $A/A_{\odot}$ is the total abundance ratio (gas+dust) in proto-Solar abundance units of \citet{Lodders2009} and $\delta_{X}$ is the depletion of the X element from the gas phase i.e., the fraction of that element locked up in dust. The abundances and depletions have been calculated using the values from the best fit model. }    
\end{table*}

\subsection{What form solid iron does take? }
\label{iron}

Identifying the composition of solid-phase iron is important to understand how dust grows in the ISM. About 90\% of the total amount of iron is missing from the gas phase and is believed to be locked up in dust grains. It has been discussed in the literature that more than 65\% of the iron is injected into the ISM in gaseous form. Therefore most of its grain growth should take place in the ISM (\citealt{Dwek2016}).
In this work, we are able to identify the solid compounds that make up the 90\% of depleted iron. 
In addition to silicates discussed in Section \ref{mine}, we find that on average 15\% of the total dust column in our sight-lines is in the form of metallic iron. 

Additional reservoirs of iron, other than silicates and metallic iron, have been extensively discussed in the literature. Ferromagnetic inclusions ($\rm Fe_{3}O_{4}$) could be possible iron carriers. In particular, dust grain inclusions of ferromagnetic material, such as magnetite, have been discussed in the view of grain alignment due to their ferromagnetic properties (\citealt{Lazarian}). Iron sulfides are other possible iron carriers. However, the amount of depleted sulfur still remains an open question. Here, we find that less than 2\% of the total dust column density in our lines of sight is in other forms, such as sulfides ($\rm FeS$ and $\rm Fe_{2}S$) or iron oxides (magnetite, $\rm Fe_{3}O_{4}$). 

Our study reveals that $\rm \sim 40\%$ of the dust-phase Fe atoms are in metallic form, and the rest $\rm \sim 60\%$ of iron is in the form of silicates. This result is an average from all the sources, calculating the mass density of iron atoms that reside in each chemical compound. Since the amount of sulfides and other Fe bearing compounds is significantly smaller, we do not include them in this calculation. \citet{Tamanai2017} discuss the distribution of Fe in presolar silicate grains originating from AGB stars and supernovae. They find that a significant fraction of iron atoms are concentrated in metallic Fe nano-particles, and not distributed in the silicate lattice. In particular, they show a wide range of Fe atom fractions in metallic Fe, with the value going from 0.2 to 1. Our value (0.4 or $\rm 40\%$) is within the limits of this study. We therefore conclude that it is possible that a significant fraction of Fe that condenses in stellar environments resides in the form of metallic iron. 

\citet{Draine_2021} discuss about the fraction of solid phase iron ($f_{Fe}$) included in metallic form. Their value of $f_{Fe}$ is considered to be small, <10\%, due to the ferromagnetic properties of metallic iron which would generate thermal magnetic-dipole emission with unusual spectral and polarization characteristics \citep{Draine_2013}. Our value of $\sim 40\%$ (on average) departs significantly from this concept. Table 2 of \citet{Draine_2021} presents the chemistry of grains composing the 'astrodust' material, showing that a large amount of the solid-phase iron is in other compounds such as sulfides or magnetite. The advantage of studying the XAFS in the Fe L-edges is that we can get a direct measure of the Fe content in dust, however we do not exclude possible biases to this result. Metallic iron is the only compound taken from the literature and therefore uncertainties in the energy calibration of the model might be possible.

\subsection{The oxygen budget problem}
\label{budget}

Accounting for the total oxygen budget in the ISM is a long standing problem. In dense environments of the ISM ($\rm n_{H}>7 \ cm^{-3}$), oxygen is missing from the gas phase at a level that cannot be explained only with its depletion into dust and molecules such as CO and $\rm H_{2}O$, and therefore additional reservoirs need to be found (\citealt{Jenkins2009}, \citealt{Whittet2010}). In our work we find on average that 10-20\% of oxygen is depleted into dust. Our oxygen depletion values are consistent with those found in \citet{Poteet}, who studied the IR absorption spectra towards the diffuse-dense cloud transition region of Zeta Oph and also found that about 21\% of oxygen is in dust. According to the authors, an amount of oxygen is still missing from the gas phase. A substantial fraction of interstellar oxygen could reside in other reservoirs, such as ices ($\rm H_{2}O$) or CO. 

In our study, the abundances of oxygen (both in gas and dust) along the different lines of sight are between 1.2-1.4 times the Solar value compared to \citet{Lodders2009}. Given the typical distances to our sources of 3-8 kpc (Table \ref{xmmdata}), we get average hydrogen density of $\rm \sim 0.1 \ cm^{-3}$ for the lines of sight studied here. This number should be used with caution however, it is merely an average over the entire spatial distribution of interstellar gas along our sight-lines. It is therefore possible that in this way, we underestimate the absolute value of the hydrogen density. Nevertheless, it is interesting to compare this value to the expected partitioning between gas and solids in \citet{Whittet2010}, who studied the oxygen reservoirs in a wide range of interstellar environments. They found that the oxygen budget problem starts to be significant for hydrogen densities above $\rm 7 \ cm^{-3}$. For the densities studied here the oxygen reservoir does not appear to be a problem. Our result confirms this scenario and therefore we conclude that in this study we do not need an additional reservoir of oxygen, such as $\rm H_{2}O$ or CO.

\subsection{Advantages of fitting the O and Fe together}
\label{advantages}

In this work we performed a simultaneous fit of the oxygen K and iron L edges and we were able to identify the chemical composition of interstellar dust in diffuse lines of sight. In Paper I, we presented a detailed demonstration of the dust laboratory experiments in the O K-edge and fitted the oxygen K edge structure alone, using \xmm data. With the new dust extinction model, we were able to disentangle the dust and gas contribution in the O K-edge but we could not distinguish among the different dust species as easily. In particular, the number of acceptable models (using the AIC criterion) in this work when we are fitting O and Fe together, is significantly smaller than in Paper I. Therefore, by fitting two edges together, we get a better constraint on the range of possible model solutions. This allows us to identify the responsible compounds and corresponding abundances and depletion values with less uncertainty.

For the reason described above, we judge the oxygen depletion measurement for Cyg X-2 in this work to be more accurate than in the value presented in our pilot study for Paper I. In Paper I, we fit the O K-edge alone and assumed that the dust chemistry is characterised by olivine only, resulting in a depletion value of 7\%. In this work, for the same sight line, we find a higher value of $18\pm2$\%. The latter oxygen depletion value has been calculated from the best fit dust minerals, not from assuming a single dust compound. We therefore conclude that by studying the two edges simultaneously we are able to break degeneracies that might be present from the fit of a single edge, and determine the dust chemistry in these diffuse regions in the Galaxy. 

In this study we were not able to get accurate constrains on the Si and Mg abundances and depletions. For this, we will need to examine the O K, Fe L, Mg K and Si K edges together, with a source that has the proper column density in order to give us visibility in all edges. Future X-ray observations could enable us to explore with even better accuracy the O and Fe content of interstellar dust, together with Si and Mg. This case will be examined in a future study.


\section{Conclusions}
\label{conclusions}

In this work we studied the dust chemistry towards diffuse lines of sight along the Galactic plane, using the soft X-ray band ($<1$~keV). We simultaneously fit the absorption around the O K-and Fe L-edges using \xmm and \chandra observations for 5 LMXBs, Cygnus X-2, Swift J1910.2-0546, XTE J1817-330, Ser X-1 and GX9+9. For the dust modelling we use the calculated extinction models in the O K-edge presented in \citet{Psaradaki} and, for the Fe L we use the models from Costantini et al. in prep. The dust models were computed from new laboratory data and include silicates, oxides, iron sulfides and metallic iron. The latter has been taken from the literature. Our main results can be summarised as follows.

\begin{itemize}
	 \item In this study we were able to constrain the dust grain chemistry in the diffuse medium. Mg-rich amorphous pyroxene dust ($\rm Mg_{0.75}Fe_{0.25}SiO_{3}$) represents the bulk of the dust chemistry in the diffuse environments of the ISM. In particular, we find an average of $\rm \sim 70\%$ of this silicate towards Cygnus X-2, Swift J1910.2-0546, XTE J1817-330 and Ser X-1. Only in the line of sight towards GX 9+9, amorphous olivine ($\rm MgFeSiO_{4}$) appears to be the dominant compound.
	 \item We find an upper limit on the silicate crystallinity of 15\%. This value is broadly consistent with the crystallinity observed in the infrared.
	 \item We find that on average 15\% of the total dust column in our lines of sight is in the form of metallic iron. 
	 \item Iron is heavily depleted from the gas phase into dust; more than 90\% of the iron is in the form of dust grains. 
	 \item 10-20\% of the neutral oxygen is depleted into dust, depending on the line of sight. We also find a slight over-abundance of oxygen, with respect to Solar. We conclude that in the lines of sight studied here we probe a diffuse medium and therefore, there is not oxygen budget problem as has been previously found for the translucent and dense ISM regions.  
	 \item We find Solar or slightly under-Solar abundances of iron. The latter could be attributed to self-shielding and the missing large grain population from this study ($\rm >0.25 \mu m$). Dust extinction by large particles will be examined in a future work.    
	 \item The simultaneous fit of the O K-and Fe L-edges gave us a stronger constraint into both the dust depletion and the chemical composition of grains in the diffuse regions of the ISM. 
	 \item Our new X-ray dust extinction cross section in the O K and Fe L-edges enabled us to understand the chemistry of interstellar dust in the diffuse lines of sight studied here. 
\end{itemize}

\begin{acknowledgements}
We thank the referee for the suggestions that helped to improve this paper. 
The authors would also like to thank J. de Plaa and J. Kaastra for their continuous help with \spex. IP thanks A. Bosman for useful discussions on oxygen reservoirs.
This research has been supported by the Netherlands Organisation
for Scientific Research (NWO) through The Innovational Research Incentives
Scheme Vidi grant 639.042.525. Portions of this work are supported by NASA’s Astrophysics Data Analysis Program, grant number 80NSSC20K0883, under the ROSES program NNH18ZDA001N. The laboratory experiment related to this project has received funding from the European Union’s Horizon 2020 research and innovation program under grant 823717-ESTEEM3.
\end{acknowledgements}

%

\vspace{-0.4cm}
\bibliographystyle{aa}
\bibliography{references}

\end{document}